\documentclass[prd,nofootinbib,english]{revtex4}

\usepackage{graphicx,float}
\usepackage{amsmath,amssymb,amsfonts}
\usepackage{mathrsfs}
\usepackage{epsfig,color}
\usepackage[thinlines]{easytable}
\usepackage{pdfpages}
\usepackage{array}
\usepackage{cancel}
\usepackage{mathtools}
\usepackage{accents}
\usepackage{subfigure}
\usepackage{enumitem}
\usepackage[dvipsnames]{xcolor}
\usepackage{hyperref}
\usepackage{verbatim}
\hypersetup{
	colorlinks=true,
	linkcolor=blue,
	filecolor=magenta,
	citecolor=blue
}

\begin{document}
\hfill  USTC-ICTS/PCFT-25-20

\title{Stability Analysis of the Possible Consistent Model of Parity Violations in the Symmetric Teleparallel Gravity: Generalized Background Solutions}

\author{Yeheng Tong}
\email{yhtong19@mail.ustc.edu.cn}
\affiliation{Interdisciplinary Center for Theoretical Study, University of Science and Technology of China, Hefei, Anhui 230026, China}

\begin{abstract}
In this paper, we consider a symmetric teleparallel gravity model that extends the general relativity equivalent model by several parity violating interactions between the gravitational field and a scalar field. We derive three different families of background solutions in flat FRW universe, with three classes of different connections. Through investigations on the linear cosmological perturbations, we show that one of the vector modes of this model will evolve into a ghost field at high energy, and the ghost instability can be cancelled only under specific combinations of the coefficients. On two of three families of backgrounds, such combination remains the same as the one we have investigated in our previous work; while on the other family of background, one additional condition should be taken into consider.
\end{abstract}

\maketitle

\section{Introduction}

In recent years, there are many interests in investigating possible parity violations in gravity theories in the literature, partly stimulated by the experimental detections of gravitational waves (GWs) \cite{ligo1,ligo2} and the developments in the cosmic microwave background radiation (CMB) experiments \cite{CMB1,CMB2}. Among the most prominent parity-violating models within Riemannian geometry is the so-called Chern-Simons (CS) modified gravity \cite{CSgravity1,CSgravity2}, which modifies general relativity (GR) by a gravitational CS term $\mathcal{L}_{CS}\sim\phi\varepsilon^{\mu\nu\rho\sigma}{R^{\alpha\beta}}_{\mu\nu}R_{\alpha\beta\rho\sigma}$, where $\phi$ is a scalar field, $R_{\mu\nu\rho\sigma}$ is the Riemann tensor constructed from Levi-Civita connection, $\varepsilon^{\mu\nu\rho\sigma}=\epsilon^{\mu\nu\rho\sigma}/\sqrt{-g}$ is the Levi-Civita tensor, $\epsilon^{\mu\nu\rho\sigma}$ is the totally antisymmetric symbol and $g$ is the determinant of the metric. A distinctive feature of CS modified gravity is the so-called amplitude birefringence phenomenon, \textit{i.e.}, an attenuation disparity in the amplitudes of the left- and right-handed polarized components of GWs, while no difference between their velocities is presented. However, the CS modified gravity suffers from the problem of vacuum instability because one of the circularly polarized components of GWs becomes a ghost mode at high frequencies \cite{CSgravity3} (characterized by a kinetic term with the wrong sign). Further extensions to the CS modified gravity were made in Refs.\cite{Crisostomi:2017ugk,Gao:2019liu,Zhao:2019xmm}, but these efforts were deemed insufficient\cite{Bartolo:2020gsh}. These results suggest that constructing a ghost-free, parity-violating gravitational theory within Riemannian geometry remains a significant challenge.

To explore consistent parity-violating gravity models, it is necessary to extend beyond the Riemannian geometry. In this context, a ghost-free model called the Nieh-Yan modified Teleparallel Gravity (NYTG) model \cite{NY1,NY2} was proposed. The NYTG model is based on the teleparallel geometry (TG) \cite{Tele1,tele2021} where the spacetime is endowed with a metric-compatible and curvature-free connection, and the gravity is identified with the torsion $\mathcal{T}^{\lambda}_{~\mu\nu}$. One can construct a GR equivalent model within the framework of TG, called Teleparallel Equivalent General Relativity (TEGR). The NYTG model \cite{NY1,NY2} modifies TEGR by the coupling between a scalar field $\phi(x)$ and the Nieh-Yan density \cite{Nieh:1981ww}  $(1/2)\varepsilon^{\mu\nu\rho\sigma}\mathcal{T}^{\lambda}_{~\mu\nu}\mathcal{T}_{\lambda\rho\sigma}$. Cosmological applications of the NYTG model\cite{NY1,NY2} reveal a velocity birefringence phenomenon, \textit{i.e.}, difference between the propagating velocities of the left- and right-handed polarized components of GWs around the Friedmann-Robertson-Walker (FRW) background, while the damping rates of their amplitudes are the same. The NYTG model also remained consistent with the results of most local tests in the solar system at the post-Newtonian order \cite{Rao:2021azn,Qiao:2021fwi}.

Besides teleparallel framework, another similar non-Riemannian framework to build gravity models is the so-called symmetric teleparallel gravity (STG) \cite{STGR}, where the spacetime is endowed with a metric and a curvature-free and torsionless connection, and gravity manifests purely through non-metricity  $Q_{\alpha\mu\nu}=\nabla_\alpha g_{\mu\nu}$. Within this framework, the Symmetric Teleparallel Equivalent of General Relativity (STEGR) serves as a GR-equivalent model. Similar to the CS modified gravity and the NYTG model, the simplest parity violating extension to STEGR involves the coupling $\phi(x)\varepsilon^{\mu\nu\rho\sigma}Q_{\mu\nu\alpha}Q_{\rho\sigma}^{~~~\alpha}$. While this modification presents velocity birefringence in cosmological tensor perturbations \cite{Conroy:2019ibo}, it introduces ghost instability for vector perturbations \cite{STPV1}. To address the instability, we proposed a generalized STG model\cite{STPV2} which extends STEGR by several parity violating interactions between the scalar field and the parity-odd terms, which are quadratic in the non-metricity tensor, \textit{i.e.} $\mathcal{L}_{PV}\sim\varepsilon QQ\hat{\nabla}\phi\hat{\nabla}\phi$. In our previous work, analysis revealed that around the simplest background, while tensor perturbations retain velocity birefringence, high-frequency ghost modes arise in vector perturbations, and a specific tuning of the parity-odd coupling coefficients can eliminate these ghosts. Further studies about this model have been published recently, \textit{e.g.}, the examination through GWTC-3 \cite{GWTC-3}.

Since in the STG framework, the solution of the system depends on both metric and the connection, more general cosmological solutions may exist beyond the one studied in our previous work\cite{STPV2}. While our earlier study focused on flat FRW metrics with specific connections, alternative connection choices could yield different background solutions where the ghost-free condition may require modification. In this paper, we will investigate extended classes of background solutions, analyze their linear perturbations, and assess whether the condition derived in our previous work remains universally valid or requires modification for these new configurations.

This paper is organized as follows. In Sec. \ref{basic knowledge}, we will review the STG model with parity violation we will consider in this paper. In Sec. \ref{cosmology}, we will apply this model to cosmology and introduce the three families of broader background solutions. In Sec. \ref{QuAcs} we present our main results about the studies on the linear cosmological perturbations, and introduce the modification of ghost-free condition. Section \ref{conclusion} is a summary.

\section{The Symmetric Teleparallel Gravity Model with Parity Violations}\label{basic knowledge}

In this paper, we will take the unit $8\pi G=1$ and the convention for the metric signature: $\left\lbrace +,-,-,-\right\rbrace$. As usual, the Greeks $\mu,\nu,\rho,\dots=0,1,2,3$ are used to represent spacetime tensor indices, and the Latins $i,j,k,\dots=1,2,3$ are used to denote the spatial components. As mentioned before, the STG theory is considered as a constrained metric-affine theory, formulated by a metric $g_{\mu\nu}$ and an affine connection ${\Gamma^\lambda}_{\mu\nu}$, which is curvature-free and torsionless:
\begin{equation}	       {R^\lambda}_{\rho\mu\nu}:=\partial_\mu{\Gamma^\lambda}_{\nu\rho}+{\Gamma^\lambda}_{\mu\sigma}{\Gamma^\sigma}_{\nu\rho}-\left\lbrace\mu\leftrightarrow\nu\right\rbrace=0~,~\mathcal{T}^\lambda_{~\mu\nu}:=\Gamma^\lambda_{~\mu\nu}-\Gamma^\lambda_{~\nu\mu}=0.
\end{equation}
With these constraints, the affine connection can be generally expressed as\cite{STPV1,STPV2} 
\begin{equation}\label{conc}
	{\Gamma^\lambda}_{\mu\nu}=\frac{\partial x^\lambda}{\partial y^\alpha}\partial_\mu\partial_\nu y^\alpha~,
\end{equation}
where the four functions $y^\alpha$ themselves form a special coordinate system in which all the components of the affine connection vanish. One can carry out the calculations by fixing to this special coordinate system. In fact, such "coincident gauge" had been frequently adopted in the literature. However, for the purpose of making our analysis as general as possible, we prefer to work in an arbitrary coordinate system in this paper. In this case, both the metric $g_{\mu\nu}$ and the four functions $y^\mu$ should be considered as the fundamental variables of the STG theory.

Within STG, the gravity is attributed to the non-metricity tensor $Q_{\alpha\mu\nu}\equiv\hat{\nabla}_\alpha g_{\mu\nu}=\partial_\alpha g_{\mu\nu}-{\Gamma^\lambda}_{\alpha\mu}g_{\lambda\nu}-{\Gamma^\lambda}_{\alpha\nu}g_{\mu\lambda}$, 
which measures the failure of the connection to be metric compatible. The STEGR model, which is equivalent to GR within the symmetric teleparallel framework, has the following action:
\begin{equation}\label{Qac}
	S_{STEGR}=\frac{1}{2}\int d^4x \sqrt{-g}\mathbb{Q} \equiv \frac{1}{2}\int d^4x \sqrt{-g} \left( \frac{1}{4}Q_{\alpha\mu\nu}Q^{\alpha\mu\nu}-\frac{1}{2}Q_{\alpha\mu\nu}Q^{\mu\nu\alpha}-\frac{1}{4}Q_{\alpha}Q^{\alpha}+\frac{1}{2}Q_{\alpha}\tilde{Q}^{\alpha}\right)~,
\end{equation}
where $Q_\alpha=Q_{\alpha\mu\nu}g^{\mu\nu}$ and $\tilde{Q}_\alpha=Q_{\rho\sigma\alpha}g^{\rho\sigma}$ are two non-metricity vectors. This action is identical to the Einstein-Hilbert action up to a boundary term: 
\begin{equation}
	S_{STEGR}=\frac{1}{2}\int d^4x \sqrt{-g} \left[-\mathring{R}-\mathring{\nabla}_{\alpha}(Q^\alpha-\bar{Q}^\alpha)\right],
\end{equation}
where both the curvature scalar $\mathring{R}$ and the covariant derivative $\mathring{\nabla}_{\alpha}$ is associated with the Levi-Civita connection. 

Similar to the CS modified gravity and the NYTG model,  a simple parity violating extension to STEGR can be realized by introducing an extra parity-odd term 
$S_{PV}\sim \int d^4x\sqrt{-g} \phi\varepsilon^{\mu\nu\rho\sigma}\,Q_{\mu\nu\alpha}\,{Q_{\rho\sigma}}^\alpha$ to the action (\ref{Qac}). 
However, as pointed out in Ref. \cite{STPV1}, with this modification, the vector perturbations in the gravity sector are promoted to be propagating dynamical modes, and one of the vector modes becomes a ghost mode at high momentum scales.

In our last work\cite{STPV2}, we turned to the more general (and of course more complex) case, where parity violating extensions in gravity are provided by several interactions between the scalar field and the parity-odd terms which are quadratic in the nonmetricity tensor. At the same time, we should guarantee the equations of motion of the metric and the scalar field to be second order. This is in order to exclude the Ostrogradsky ghosts, which are originated from higher derivatives. Totally there are seven couplings satisfying the requirements, as listed below\cite{STPV2,Conroy:2019ibo},
\begin{equation}
	\begin{aligned}\label{PVc}
		M_1&=\varepsilon^{\mu\nu\rho\sigma}\,Q_{\mu\nu\alpha}\,{Q_{\rho\sigma}}^\alpha\,\nabla_\beta \phi\,\nabla^\beta \phi,\\
		M_2&=\varepsilon^{\mu\nu\rho\sigma}\,Q_{\mu\nu\alpha}\,{Q_{\rho\sigma}}^\beta\,\nabla^\alpha \phi\,\nabla_\beta \phi,\\
		M_3&=\varepsilon^{\mu\nu\rho\sigma}\,Q_{\mu\nu\alpha}\,{Q_{\rho}}^{\alpha\beta}\,\nabla_\sigma \phi\,\nabla_\beta \phi,\\
		M_4&=\varepsilon^{\mu\nu\rho\sigma}\,Q_{\mu\nu\alpha}\,{Q^{\alpha\beta}}_\rho\,\nabla_\sigma \phi\,\nabla_\beta \phi,\\
		M_5&=\varepsilon^{\mu\nu\rho\sigma}\,Q_{\mu\nu\alpha}\,{Q^{\beta\alpha}}_\rho\,\nabla_\sigma \phi\,\nabla_\beta \phi,\\
		M_6&=\varepsilon^{\mu\nu\rho\sigma}\,Q_{\mu\nu\alpha}\,Q_{\rho}\,\nabla^\alpha\phi\,\nabla_\sigma \phi,\\
		M_7&=\varepsilon^{\mu\nu\rho\sigma}\,Q_{\mu\nu\alpha}\,\bar{Q}_{\rho}\,\nabla^\alpha\phi\,\nabla_\sigma \phi.
	\end{aligned}
\end{equation}
Then the action of parity violation is written as
\begin{equation}\label{PVa}
	S_{PV}=\int d^4x\sqrt{-g}\sum_{a} c_a(\phi,\nabla^\mu \phi \nabla_\mu \phi) M_a\equiv\sum_{a} S_{PVa}~,
\end{equation}
with $a=1,2,\cdots,7$. The coupling coefficients $c_a$ can also rely on the scalar field and its first derivatives. 

However, the parity violating terms list above are not linearly independent. The two following identities hold\cite{Chen:2022wtz}:
\begin{equation}\label{PVId}
    M_2 -M_3 +M_4=M_6 -M_7,~\frac{1}{2}M_1=M_3 -M_5.
\end{equation}
So only five of the terms need to be considered. In this paper, we remove $M_3$ and $M_7$ and retain the other terms, so the full action of the model under consideration is 
\begin{equation}\label{S2}
	S=S_{STEGR}+S_{PV}+S_\phi= \int d^4x\sqrt{-g}\left[\frac{\mathbb{Q}}{2}+\sum_{a} c_a(\phi,\nabla^\mu \phi \nabla_\mu \phi) M_a +\frac{1}{2} g^{\mu\nu}\partial_\mu\phi\partial_\nu\phi-V(\phi)\right]~,
\end{equation}
where we have neglected the matter other than the scalar field $\phi$ in the Universe, $a=1,2,4,5,6$. Performing the variational principle on the action (\ref{S2}) yields the equations of motion (EoMs) of the model: 
\begin{align}
    \mathring{G}^{\mu\nu}+\sum_{a} N_a^{\mu\nu}&=T^{\mu\nu},\label{PVmeteq}\\
    \sum_{a}\nabla_\mu\nabla_\nu\left(c_a \sqrt{-g}{\mathbb{P}_a}^{(\mu\nu)}_{\quad~\alpha}\right)&=0,\label{PVconeq}
\end{align}
where
\begin{equation}
    N_a^{\mu\nu}=-\frac{2}{\sqrt{-g}} \frac{\delta S_{PVa}}{\delta g_{\mu\nu}}
\end{equation}
are the modifications to the EoMs of metric arising from the parity violating terms, and 
\begin{equation}
    \mathbb{P}_a^{\alpha\mu\nu}:=\frac{1}{2\sqrt{-g}} \frac{\partial \left(\sqrt{-g}M_a\right)}{\partial Q_{\alpha\mu\nu}}
\end{equation}
are hyper-potentials of the parity violating terms. The absence of terms derived from $\mathbb{Q}$ in EoMs of connection (\ref{PVconeq}) stems from the Bianchi identity in STEGR\cite{TePa}; The matter sector does not contribute to the EoMs of connection either, because the matter fields exhibit minimal coupling to the gravitational sector.

Our preceding analysis\cite{STPV2} has demonstrated a preliminary ghost-free condition based on the simplest background, which is $b_1\equiv 2c_1+2c_2-c_4-c_5-c_7=0$. However, it remains open to discussion whether this condition pertains to other backgrounds, which will be discussed in this paper. In this paper, since terms $M_3$ and $M_7$ are neglected, the ghost-free condition also reduces to
\begin{equation}\label{noghcon}
    b_1\equiv 2c_1+2c_2-c_4-c_5=0.
\end{equation}
We shall assume this condition to always hold in subsequent calculations.

\section{Application to Cosmology: More Background Solutions}\label{cosmology}
\subsection{The background dynamics}

For simplicity, we take the spatially flat FRW universe as the background, whose line element is given by 
\begin{equation}
	ds^2=a^2(\eta)(d\eta^2-\delta_{ij}dx^i dx^j),
\end{equation}
where $\eta$ is the conformal time. In GR we only need the metric to define a background, since only the metric is the fundamental variable; while in the STG theory, the four functions $y^\mu$ determining affine connection are also considered as fundamental variables. Hence, besides the metric, these four functions should also be given. In our previous work\cite{STPV2}, the simplest background solution of the affine connection $\bar{\Gamma}^{\lambda}_{~\mu\nu}=0$ has been considered, which means to set $y^\mu =x^\mu\equiv\left\lbrace\eta,x^i \right\rbrace$ for the cosmic background. 
While in this paper, we will consider more general background solutions.

The cosmological principle rigorously constrains the spacetime metric to the FRW form. However, since metric and affine connection are independent in the symmetric teleparallel framework, it is an opening question whether the affine connection is restricted by the cosmological principle at the same time. Nevertheless, it remains reasonable 
to require that the affine connection respects the cosmological principle, which lead to a vanishing Lie derivative of the non-metricty tensor with respect to the Killing vectors $\varsigma$ of flat FRW metric: 
\begin{equation}
    \mathscr{L}_\varsigma Q_{\alpha\mu\nu}=0.
\end{equation}
And this condition constrains the non-metricity tensor to adopt the following specific functional form\cite{Iosifidis:2020gth}:
\begin{equation}\label{sy-ic-Q} 
    Q_{\alpha\mu\nu}= \mathcal{A}(\eta)g_{\mu\nu}U_{\alpha} +\mathcal{B}(\eta) \left(g_{\alpha\mu}U_\nu +g_{\alpha\nu}U_\mu\right) +\mathcal{C}(\eta)U_\mu U_\nu U_\alpha,
\end{equation}
where $\mathcal{A}$, $\mathcal{B}$ and $\mathcal{C}$ are three smooth functions of time, and $U_\mu$ is a four-velocity co-vector. When adopting the flat FRW metric as the background spacetime, the general non-metricity tensor structure reduces to
\begin{equation}\label{sy-ic-Qb} 
    \bar{Q}_{\alpha\mu\nu}=a^2\left[ \mathcal{A}(\eta)\eta_{\mu\nu}U_{\alpha} +\mathcal{B}(\eta)\left(\eta_{\alpha\mu}U_\nu +\eta_{\alpha\nu}U_\mu\right)\right] +\mathcal{C}(\eta)U_\mu U_\nu U_\alpha,
\end{equation}
where the functions $\mathcal{A}$, $\mathcal{B}$ and $\mathcal{C}$, along with the four-velocity $U_\mu$ remain to be determined through the EoMs of affine connection.

In the model (\ref{S2}), the hyper-potentials from the parity violating terms are
\begin{equation}
    \begin{aligned}
        {\mathbb{P}_1}^{\mu\nu}_{\quad\lambda} &= 2\varepsilon^{\mu\nu\rho\sigma}Q_{\rho\sigma\lambda}\nabla_\beta \phi \nabla^\beta \phi,\\
        {\mathbb{P}_2}^{\mu\nu}_{\quad\lambda} &= 2\varepsilon^{\mu\nu\rho\sigma}Q_{\rho\sigma}^{~~~\beta}\nabla_\beta \phi \nabla_\lambda \phi,\\
        {\mathbb{P}_4}^{\mu\nu}_{\quad\lambda} &= \varepsilon^{\mu\nu\rho\sigma} Q_{\lambda\rho}^{~~~\beta}\nabla_\beta \phi \nabla_\sigma \phi +\varepsilon^{\rho\sigma\alpha}_{\quad~\lambda} Q_{\rho\sigma}^{~~~\mu}g^{\nu\beta}\nabla_\alpha \phi \nabla_\beta \phi,\\
        {\mathbb{P}_5}^{\mu\nu}_{\quad\lambda} &= \varepsilon^{\mu\nu\rho\sigma} {Q^\beta}_{\lambda\rho}\nabla_\beta \phi \nabla_\sigma \phi +\varepsilon^{\rho\sigma\alpha}_{\quad~\lambda} Q_{\rho\sigma}^{~~~\mu}g^{\nu\beta}\nabla_\alpha \phi \nabla_\beta \phi,\\
        {\mathbb{P}_6}^{\mu\nu}_{\quad\lambda} &= \varepsilon^{\mu\nu\rho\sigma} Q_{\rho}\nabla_\lambda \phi \nabla_\sigma \phi -\varepsilon^{\rho\sigma\alpha\mu}\delta^\nu_\lambda Q_{\rho\sigma}^{~~~\beta}\nabla_\alpha \phi \nabla_\beta \phi.
    \end{aligned}
\end{equation}
Substituting Eq. (\ref{sy-ic-Q}) into the aforementioned five equations, summing them, and subsequently symmetrizing with respect to the indices $\mu$ and $\nu$, the EoMs of connection on flat FRW background are obtained:
\begin{equation}
        \sum_a  c_a{\mathbb{P}_a}^{(\mu\nu)}_{\quad~\lambda} = \left[\left(c_4+c_5\right) \varepsilon^{\rho\sigma~~(\mu}_{~~~\lambda}g^{\nu)\alpha} -c_6 \varepsilon^{\rho\sigma\alpha(\mu} \delta^{\nu)}_\lambda\right] \left(\mathcal{A}-\mathcal{B}\right)U_\rho \nabla_\alpha \phi \nabla_\sigma \phi,
\end{equation}
It is obvious that, as long as $U_\rho\propto\nabla_\rho \phi$, the summation of the hyper-potentials vanishes automatically:
\begin{equation*}
    \sum_a {\mathbb{P}_a}^{(\mu\nu)}_{\quad~\lambda}=0.
\end{equation*}
So the EoMs of connection (\ref{PVconeq}) are also satisfied. Noticing that on the background, the scalar field is $\bar{\phi}=\bar{\phi}(\eta)$, so we have $\nabla_\rho \bar{\phi}=\{\bar{\phi}',0,0,0\}$, which leads to $U_\rho=\{a,0,0,0\}$. With these notifications, the background non-metricity tensor is
\begin{equation}\label{Qbg} 
    \bar{Q}_{\alpha\mu\nu}=a^2\left[ 2\mathcal{A}(\eta)\eta_{\mu\nu}d_{\alpha} +\mathcal{B}(\eta)\left(\eta_{\alpha\mu}d_\nu +\eta_{\alpha\nu}d_\mu\right) +2\mathcal{C}(\eta)d_\mu d_\nu d_\alpha\right],
\end{equation}
where $d_\alpha:=\{1,0,0,0\}$, and functions $\mathcal{A}$, $\mathcal{B}$ and $\mathcal{C}$ have been redefined.

Now we have come up with a background non-metricity tensor (\ref{Qbg}) that automatically satisfies EoMs of connection (\ref{PVconeq}). However, it still need to be examined whether the affine connection associated with (\ref{Qbg}) is curvature-free and torsionless. The affine connection associated with (\ref{Qbg}) can be derived with the following formula:
\begin{equation}
    {\Gamma^\lambda}_{\mu\nu}=\mathring{\Gamma}^\lambda_{~~\mu\nu}+\frac{1}{2}\left( {Q^\lambda}_{\mu\nu} - {Q_{\mu\nu}}^\lambda -{Q_{\nu\mu}}^\lambda \right).
\end{equation}
And the result is
\begin{equation}\label{bgGamma}
    \bar{\Gamma}^{0}_{~~00}=\mathcal{H}-\mathcal{A}-\mathcal{B}-\mathcal{C},\bar{\Gamma}^{0}_{~~ij}=\left(\mathcal{H}-\mathcal{A}+\mathcal{B}\right)\delta_{ij},\bar{\Gamma}^{l}_{~~0i}=\bar{\Gamma}^{l}_{~~i0}=\left(\mathcal{H}-\mathcal{A}\right)\delta^{l}_{i},
\end{equation}
while all the unlisted components vanish. It is obvious that (\ref{bgGamma}) is torsionless, but further calculation is needed to examine whether it is curvature-free. The curvature of (\ref{bgGamma}) can be derived directly:
\begin{equation}\label{bgR}
    \begin{aligned}
        \bar{R}^0_{~~k0i}&=\left[\left(\mathcal{H}-\mathcal{A}+\mathcal{B}\right)' -\left(\mathcal{H}-\mathcal{A}+\mathcal{B}\right)\left(\mathcal{B}+\mathcal{C}\right)\right]\delta_{ik},\\
        \bar{R}^l_{~~00i}&=\left[\left(\mathcal{H}-\mathcal{A}\right)' +\left(\mathcal{H}-\mathcal{A}\right) \left(\mathcal{B}+\mathcal{C}\right) \right] \delta^l_i,\\
        \bar{R}^l_{~~kij}&=\left(\mathcal{H}-\mathcal{A}+\mathcal{B}\right) \left(\mathcal{H}-\mathcal{A}\right) \left(\delta^l_i \delta_{jk} -\delta^l_j\delta_{ik}\right),
    \end{aligned}
\end{equation}
while all the unlisted components vanish. As the curvature in symmetric teleparallel framework, the components in (\ref{bgR}) should also vanish, which yields the following equations:
\begin{align}
    \left(\mathcal{H}-\mathcal{A}+\mathcal{B}\right)' -\left(\mathcal{H}-\mathcal{A}+\mathcal{B}\right)\left(\mathcal{B}+\mathcal{C}\right)&=0,\label{Req1}\\
    \left(\mathcal{H}-\mathcal{A}\right)' +\left(\mathcal{H}-\mathcal{A}\right)\left(\mathcal{B}+\mathcal{C}\right)&=0,\label{Req2}\\
    \left(\mathcal{H}-\mathcal{A}+\mathcal{B}\right)\left(\mathcal{H}-\mathcal{A}\right)&=0.\label{Req3}
\end{align}
The equations above admit three distinct families of solutions, which can be categorized as follows: when $\mathcal{B}=0$, 
\begin{enumerate}
    \item Degenerate Family: $\mathcal{A}=\mathcal{H}$ and $\mathcal{C}=F$, so that
    \begin{equation}\label{Sol3}
        \bar{Q}_{\alpha\mu\nu}= 2a^2\left(\mathcal{H} \eta_{\mu\nu} d_{\alpha} +F d_\mu d_\nu d_\alpha\right);
    \end{equation}
\end{enumerate}
and when $\mathcal{B}\neq 0$,
\begin{enumerate}[resume]
    \item Non-degenerate Family 1: $\mathcal{A}=\mathcal{H}+F$, $B=F$ and $\mathcal{C}=-F'/F-F$, so that
    \begin{equation}\label{Sol1}
        \bar{Q}_{\alpha\mu\nu}= a^2\left[2\left(\mathcal{H}+F\right)\eta_{\mu\nu} d_{\alpha} +F\left(\eta_{\alpha\mu} d_\nu +\eta_{\alpha\nu} d_\mu\right) -2\left(\frac{F'}{F}+F\right) d_\mu d_\nu d_\alpha\right];
    \end{equation}
     \item Non-degenerate Family 2: $\mathcal{A}=\mathcal{H}$, $B=F$, and $\mathcal{C}=F'/F-F$, so that
    \begin{equation}\label{Sol2}
        \bar{Q}_{\alpha\mu\nu}= a^2\left[2\mathcal{H}\eta_{\mu\nu} d_{\alpha} +F\left(\eta_{\alpha\mu} d_\nu +\eta_{\alpha\nu} d_\mu\right) +2\left(\frac{F'}{F}-F\right) d_\mu d_\nu d_\alpha\right].
    \end{equation}
\end{enumerate}
In these three families of solutions, the specific functional form of the smooth function $F=F(\eta)$ determines the distinct members within each family of solutions, leading to physically and mathematically differentiated cases. Specifically, the simplest background solution $\bar{\Gamma}^\lambda_{~~\mu\nu}=0$ discussed in our previous work\cite{STPV2} precisely belongs to the degenerate solution family (\ref{Sol3}), and corresponds to the case $F=0$. Furthermore, taking the limit $F\rightarrow 0$ while requiring $F'/F$ to remain bounded causes the non-degenerate solutions 1 (\ref{Sol1}) and 2 (\ref{Sol2}) to reduce to the degenerate solution (\ref{Sol3}).

After some calculations, one can straight forwardly find that the extra parity violating terms in the model (\ref{S2}) do not have effects on the evolution of the background. Thus the background dynamics is the same as in GR, 
\begin{equation}
	3\mathcal{H}^2=a^2\rho_\phi,\quad-2\mathcal{H}'-\mathcal{H}^2=a^2 p_\phi,\quad\phi''+2\mathcal{H}\phi'+a^2\phi=0,
\end{equation}
where the prime represents the derivative with respect to the conformal time $\eta$ and $\mathcal{H}=a'/a$ is the conformal Hubble rate. As usual, $\rho_\phi= \phi'^2/(2a^2)+V(\phi)$ and $p_\phi=\phi'^2/(2a^2)-V(\phi)$ are the energy density and pressure of the scalar field, respectively. This result holds universally across all three families of background solutions (\ref{Sol3},\ref{Sol1},\ref{Sol2}).

\section{Application to Cosmology: Perturbation Analysis}\label{QuAcs}
\subsection{Cosmological perturbations}

After the background solutions (\ref{Sol3},\ref{Sol1},\ref{Sol2}) determined, we then consider the linear cosmological perturbations of the model (\ref{S2}). With the Scalar-Vector-Tensor decomposition, the perturbed metric is contained in the following parametrized line element:
\begin{equation}
	ds^2=a^2\left\lbrace (1+2A)d\eta^2+2(\partial_i B+B_i)d\eta dx^i-\left[(1-2\psi)\delta_{ij}+2\partial_i\partial_jE+\partial_jE_i+\partial_i E_j+h_{ij}\right]dx^i dx^j \right\rbrace~,
\end{equation}
where $A,\psi,B,E$ are the scalar perturbations, $E_i, B_i$ are the vector perturbations which satisfy the transverse conditions: $\partial_i B_i=\partial_i E_i=0$, and the tensor perturbations $h_{ij}$ are transverse and traceless: $\partial_i h_{ij}=0$ and $\delta^{ij}h_{ij}=0$.

Unlike the case we considered in our last work\cite{STPV2}, the backgrounds we presented in the last subsection (\ref{Sol3},\ref{Sol1},\ref{Sol2}) are with non-zero connections, which results in more complicated expressions of perturbed connections. Considering the perturbations of the four functions $y^\alpha=\bar{y}^\alpha +\delta y^\alpha$, and define
\begin{equation}
    u^\lambda=\frac{\partial x^\lambda}{\partial \bar{y}^\alpha} \delta y^\alpha.
\end{equation}
Then the perturbed connection can be built from the expansion of Eq.(\ref{conc}). After some tedious calculations, the result up to the second order can be written as (see the appendix for details)
\begin{equation}\label{percon}
	{\Gamma^\lambda}_{\mu\nu}= \bar{\Gamma}^{\lambda}_{~~\mu\nu} +\bar{\nabla}_\mu \bar{\nabla}_\nu u^\lambda -\bar{\nabla}_\mu \bar{\nabla}_\nu u^\rho \bar{\nabla}_\rho u^\lambda,
\end{equation}
where $\bar{\nabla}$ refers to the covariant derivative associated with connection $\bar{\Gamma}^{\lambda}_{~~\mu\nu}$. Furthermore, the perturbation $u^\lambda$ is decomposed as  $u^\lambda\equiv\left\lbrace u^0,\partial_i u+u_i \right\rbrace$ with the transverse condition: $\partial_i u_i=0$. So we have two more scalar perturbations $u^0$ and $u$, and one more vector perturbation $u_i$. And as usual, the background and perturbation decomposition for the scalar field is: $\phi(\eta,\vec{x})=\phi(\eta)+\delta\phi(\eta,\vec{x})$.

We shall then introduce the gauge transformation of perturbation $u^\lambda$. The diffeomorphism of a connection in STG framework generated by infinitesimal vector $\xi$ can be written as
\begin{equation}
    \delta_\xi {\Gamma^\lambda}_{\mu\nu}= -\nabla_\mu \nabla_\nu \xi^\lambda,
\end{equation}
since both torsion and curvature vanish in STG framework. Comparing the equation above with Eq.(\ref{percon}), the diffeomorphism of $u^\lambda$ shows up immediately:
\begin{equation}
    u^\lambda \rightarrow u^\lambda -\xi^\lambda.
\end{equation}
This result yields a new gauge choice in parallel with ones in the usual perturbation theory (\textit{e.g.}, conformal Newtonian gauge and Unitary gauge): to set
\begin{equation}
    u^0=u=0,~u_i=0.
\end{equation}
In this paper we call it "coincident gauge on the perturbation level". Since in the case $\bar{\Gamma}^{\lambda}_{~~\mu\nu}\neq 0$, the expansion of Eq.(\ref{percon}) is quite complicated, we will choose "coincident gauge on the perturbation level" to simplify the calculation. We shall emphasize that under such a gauge choice, the perturbations $E$, $\delta\phi$ together with $E_i$ should not be neglected in the calculation.

So far all the perturbations in our model have been defined, and one can derive the linear perturbation equations of the scalar, vector and tensor perturbations, respectively, by substituting these perturbation parameters to the EoMs of model (\ref{PVmeteq},\ref{PVconeq}). For our purpose in this paper, it is better to consider directly the quadratic actions for the perturbations. The linear perturbation equations can be obtained from the quadratic actions through the variational principle, so all the properties of the perturbation equations have been contained in the quadratic actions already. Furthermore, the quadratic action has an advantage of showing clearly whether there are dangerous modes in the spectrum. Hence, in this section, we will focus on the quadratic actions for the perturbations of the model (\ref{S2}).

It can be verified that on each of the three families of backgrounds (\ref{Sol3},\ref{Sol1},\ref{Sol2}), the coupling among the scalar, vector and tensor perturbations does not exist. So it is feasible to consider the quadratic actions for scalar, vector and tensor perturbations separately. The results are listed below by background:

\subsection{Quadratic actions on the degenerate backgrounds}

\subsubsection{Scalar perturbations}

For the scalar perturbations, it is not difficult to find that all the parity violating terms in Eq. (\ref{PVc}) vanish up to the second order. So, they are at least third order quantities and have no contribution to the quadratic action. Therefore, the quadratic action for the scalar perturbations of the model (\ref{S2}) is the same as the one in GR with a minimally coupled scalar field:
\begin{equation}
	S^{(2)}_S= \int d^4x \frac{a^2\bar{\phi}^{\prime 2}}{2\mathcal{H}^2}\left(\zeta'^2-\partial_i\zeta\partial_i\zeta\right)~,
\end{equation}
where $\zeta=-(\psi+\mathcal{H}\delta\phi/\phi')$ is the Mukkhanov-Sasaki variable, which is gauge invariant and denotes the curvature perturbation of the hypersurface with the constant $\phi$ field. Since the parity violating terms do not contribute to the quadratic action, one can always choose a different gauge (i.e., unitary gauge) before the calculation, while $u^0$ and $u$ are allowed to be neglected in the $S_{STEGR}$ sector. This conclusion is the same as the one in our last work\cite{STPV2}.

\subsubsection{Tensor perturbations}

For tensor perturbations, only the first and the fifth parity violating terms of Eq. (\ref{PVc}) have non-zero contributions to the quadratic action:  
\begin{equation}
	S_{PV1}=-\int d^4x\,2c_1\,\phi'^2 \epsilon_{ijk}h_{jl,i}h'_{kl}~,~ S_{PV5}=\int d^4x\,c_5\,\phi'^2 \epsilon_{ijk}h_{jl,i}h'_{kl}~,
\end{equation}
where $\epsilon_{ijk}$ is the 3-dimensional anti-symmetric symbol and $\epsilon_{123}=-1$. So the full quadratic action for tensor perturbations of the model (\ref{S2}) is
\begin{equation}\label{qSt3}
	S^{(2)}_T= \int d^4x \left\lbrace \frac{a^2}{8}\left( h'_{ij}h'_{ij} -h_{ij,k} h_{ij,k}\right) +\frac{1}{2}\left(b_2\bar{\phi}^{\prime2}\right)' \epsilon_{ijk}h_{jl,i}h_{kl}\right\rbrace~,
\end{equation}
where $b_2\equiv2c_1-c_5$ for short.

The parity violations can be shown more clearly in the Fourier space. For this, we expand the tensor perturbations $h_{ij}$ in terms of the plane wave and the circular polarization bases $e^A_{ij}$:
\begin{equation}
	h_{ij}=\sum_{A=L,R}\int d^3\vec{k}\,e^{-i\vec{k}\cdot\vec{x}}\,h^A(\eta,\vec{k})\,e^A_{ij}~,
\end{equation}
where the index $A=L,R$ denotes the left- and right-handed polarizations, the bases $e^A_{ij}$ satisfy the relations $e^A_{ij}e^{B*}_{ij}=\delta^{AB}$, and $i\epsilon_{ijk}k_i e^A_{jl}=\lambda_A k e^A_{kl}$. 
The parameter $\lambda_A=\mp 1$ for $A=L, R$, respectively, and is used to remind the helicity dependence. Then the quadratic action (\ref{qSt3}) is rewritten as
\begin{equation}\label{qStt3}
	S^{(2)}_T=\sum_{A=L,R} \int d\eta\, d^3\vec{k}\,\frac{a^2}{8} \left(h'^A h'^{A*}-\omega^{2}_{AT}h^A h^{A*}\right)~,
\end{equation}
with
\begin{equation}\label{drT3}
	\omega^2_{AT}=k^2\left[ 1+\frac{4\lambda_A}{a^2k} \left(b_2\bar{\phi}^{\prime2}\right)'\right]~.
\end{equation}
The quadratic action (\ref{qStt3}) shows no ghost mode in the tensor perturbations because both the left- and right-handed polarization modes of GWs have the right sign ($a^2>0$) in their kinetic terms. But the dispersion relation (\ref{drT3}) is helicity dependent, which implies a velocity difference between the two circular polarization modes, i.e., the velocity birefringence phenomenon.  
Both polarization modes have propagating velocities different from the speed of light, so are constrained by the event GW170817 observed by LIGO/Virgo \cite{ligo2}; this in turn puts constraints on the coefficients $c_1$ and $c_5$ of this model \cite{Conroy:2019ibo}. 

\subsubsection{Vector perturbations}

After some tedious calculations, we find the following four of five parity violating terms have contributions to the quadratic action for the vector perturbations: 
\begin{equation}
\begin{aligned}
	S_{PV1}&=\int d^4x\, 2c_1 \bar{\phi}^{\prime2} \epsilon_{ijk}\left( B_{j,i}B'_k -F B_{j,i}B_k -E_{j,il}E'_{k,l} \right),\\
	S_{PV2}&=\int d^4x\, 2c_2 \bar{\phi}^{\prime2} \epsilon_{ijk}\left(B_{j,i}B'_k -FB_{j,i}B_k\right), \\
	S_{PV4}&=-\int d^4x\, c_4 \bar{\phi}^{\prime2} \epsilon_{ijk}\left( B_{j,i}B'_k -FB_{j,i}B_k -E_{j,il}B_{k,l}\right), \\
	S_{PV5}&=\int d^4x\, c_5 \bar{\phi}^{\prime2} \epsilon_{ijk}\left(E_{j,il}E'_{k,l} -B_{j,i}B'_k +FB_{j,i}B_k\right),
\end{aligned}
\end{equation}
There are some difference from the result in our last work: additional terms with function $F$ shows up. However, these terms do not contribute to the quadratic action since $b_1=0$. Summing terms above with vector sector of $S_{STEGR}$, we have the full quadratic action for the vector perturbations of the model (\ref{S2}): 
\begin{equation}\label{qSv3}
	S^{(2)}_V=\int d^4x\, \left[ \frac{1}{4}a^2(B_{i,j}+E'_{i,j})(B_{i,j}+E'_{i,j}) +\epsilon_{ijk}\bar{\phi}^{\prime2}\left( b_1 B'_k B_{j,i} -Fb_1 B_{j,i}B_k +c_4 B_{k,l} E_{j,il} -b_2 E'_{k,l} E_{j,il}\right) \right]~.
\end{equation}

It is clear from above quadratic action that the variables $B_i$ are not dynamical modes, the variation of (\ref{qSv3}) with respect to $B_i$ yields the following constraint equation:
\begin{equation}
	-\frac{1}{2}a^2\nabla^2(B_k+E'_k) +\epsilon_{ijk}c_4 \bar{\phi}^{\prime2} \nabla^2 E_{j,i}=0.
\end{equation}
where $\nabla^2$ is the Laplacian, and $b_1=0$ has been set. It is better to solve this equation in the Fourier space. Similar to what we have done in the previous subsection for the tensor perturbations, one can expand the perturbations $B_i$ and $E_i$ in terms of the plane wave and the circular polarization bases $e^A_i$ with $A=L,R$ as
\begin{equation}\label{constraint3}
	B_i\equiv\sum_{A=L,R} \int d^3\vec{k}\,e^{-i\vec{k}\cdot\vec{x}}\,B^A(\eta,\vec{k})\,e^A_i~,
\end{equation}
and a same expansion for $E_i$. Similarly, the circular polarization bases for vector perturbations $e^A_i$ satisfy the relations $e^A_i e^{B*}_i=\delta^{AB}$ and $ik_i e^A_j \epsilon_{ijk}=k\lambda^A e^A_k$. Again, the parameters $\lambda^A=\mp 1$ for $A=L, R$ remind the helicity dependence when they appear.  
With these equipments, the constraint equation (\ref{constraint3}) is rewritten in the following form:
\begin{equation}
	a^2k^2 (B^A+E'^A)+2\lambda^A c_4 \bar{\phi}^{\prime2} k^3 E^A=0~,
\end{equation}
and can be solved as
\begin{equation}\label{BA3}
	B^A= -2a^{-2} k\lambda^A c_4\bar{\phi}^{\prime2} E^A -E'^A~.
\end{equation}
At the same time, we also rewrite the quadratic action (\ref{qSv3}) in the Fourier space:
\begin{equation}\label{qSvt3}
	S^{(2)}_V= \sum_{A=L,R} \int d\eta\, d^3\vec{k}\, \left\lbrace \frac{1}{4} a^2 k^2(B^A+E'^A)(B^{*A}+E'^{*A}) +\frac{1}{2}\lambda^A k^3 \left[ -2c_4\bar{\phi}^{\prime2} E^A B^{*A}+\left(b_2\bar{\phi}^{\prime2}\right)' E^A E^{*A}\right] \right\rbrace.
\end{equation}
Then substitute the solution (\ref{BA3}) back into the action (\ref{qSvt3}), one finally obtains the quadratic action for the vector perturbations: 
\begin{equation}\label{BV3}
	S^{(2)}_V= -\sum_{A=L,R} \int d\eta\, d^3k \, w_A^2 E^A E^{*A}~,
\end{equation}
where
\begin{equation}\label{pc3}
w_A^2= \frac{k^4}{a^2}c_4^2 \bar{\phi}^{\prime4} +\frac{1}{2}\lambda^A k^3 \left[\left(b_2-c_4\right) \bar{\phi}^{\prime2}\right]'.
\end{equation}
It is evidently an action for non-dynamical fields, and free of ghost instability of course. 

So far we have derived the whole linear perturbation theory on the background (\ref{Sol3}). We found that the results are the same as ones on the simplest background we had considered in our last work\cite{STPV2}. Such convergence behavior is reasonable, since the simplest background is just the special case $F=0$ in the degenerate family(\ref{Sol3}).

\subsection{Quadratic actions on the non-degenerate backgrounds 1}
\subsubsection{Scalar perturbations}
For scalar perturbations, same as the case in the last subsection, the parity violating terms have no contribution to the quadratic action, so the quadratic action for the scalar perturbations of the model (\ref{S2}) is also the same as the one in GR with a minimally coupled scalar field:
\begin{equation}
	S^{(2)}_S= \int d^4x \frac{a^2\bar{\phi}^{\prime 2}}{2\mathcal{H}^2}\left(\zeta'^2-\partial_i\zeta\partial_i\zeta\right)~,
\end{equation}
where $\zeta=-(\psi+\mathcal{H}\delta\phi/\phi')$ is the Mukkhanov-Sasaki variable. Similarly, this quadratic action is free of effect of the perturbations $u^0$ and $u$ from the connection.

\subsubsection{Tensor perturbations}
The calculation yields the same result with the one on the degenarate background for tensor perturbations:
\begin{equation}\label{qSt1}
	S^{(2)}_T= \int d^4x \left\lbrace \frac{a^2}{8}\left( h'_{ij}h'_{ij} -h_{ij,k} h_{ij,k}\right) +\frac{1}{2}\left(b_2\bar{\phi}^{\prime2}\right)' \epsilon_{ijk}h_{jl,i}h_{kl}\right\rbrace~.
\end{equation}

Then rewrite the effective action (\ref{qSt1}) in the Fourier space:
\begin{equation}\label{qStt1}
	S^{(2)}_T=\sum_{A=L,R} \int d\eta\, d^3\vec{k}\,\frac{a^2}{8} \left(h'^A h'^{A*}-\omega^{2}_{AT}h^A h^{A*}\right)~,
\end{equation}
with
\begin{equation}\label{drT1}
	\omega^2_{AT}=k^2\left[ 1+\frac{4\lambda_A}{a^2k} \left(b_2\bar{\phi}^{\prime2}\right)'\right]~.
\end{equation}
The same as the degenerate case, the quadratic action (\ref{qStt1}) shows no ghost mode in the tensor perturbations because both the two modes of GWs have the right sign ($a^2>0$) in their kinetic terms. On the other hand, the dispersion relation (\ref{drT1}) is helicity dependent, which implies a velocity difference between the two circular polarization modes, i.e., the velocity birefringence phenomenon. 

\subsubsection{Vector perturbations}
After some calculations, one can obtain the parity violating terms that contribute to the quadratic action for the vector perturbations: 
\begin{equation}
\begin{aligned}
	S_{PV1}&=\int d^4x\, 2c_1 \bar{\phi}^{\prime2} \epsilon_{ijk}\left( B_{j,i}B'_k +\mathcal{C} B_{j,i}B_k -E_{j,il}E'_{k,l} \right),\\
	S_{PV2}&=\int d^4x\, 2c_2 \bar{\phi}^{\prime2} \epsilon_{ijk}\left(B_{j,i}B'_k +\mathcal{C}B_{j,i}B_k\right), \\
	S_{PV4}&=-\int d^4x\, c_4 \bar{\phi}^{\prime2} \epsilon_{ijk}\left( B_{j,i}B'_k +\mathcal{C}B_{j,i}B_k -E_{j,il}B_{k,l}\right), \\
	S_{PV5}&=\int d^4x\, c_5 \bar{\phi}^{\prime2} \epsilon_{ijk}\left(E_{j,il}E'_{k,l} -B_{j,i}B'_k -\mathcal{C}B_{j,i}B_k\right),
\end{aligned}
\end{equation}
where $\mathcal{C}=-(F'/F+F)$ indicates the effect of function $F$. Similarly, these terms do not contribute to the quadratic action since $b_1=0$. Summing terms above with vector sector of $S_{STEGR}$, we have the full quadratic action for the vector perturbations of the model (\ref{S2}): 
\begin{equation}\label{qSv1}
	S^{(2)}_V=\int d^4x\, \left[ \frac{1}{4}a^2(B_{i,j}+E'_{i,j})(B_{i,j}+E'_{i,j}) +\epsilon_{ijk}\bar{\phi}^{\prime2}\left( b_1 B'_k B_{j,i} +b_1\mathcal{C} B_{j,i}B_k +c_4 B_{k,l} E_{j,il} -b_2 E'_{k,l} E_{j,il}\right) \right]~.
\end{equation}
This action is similar to Eq.(\ref{qSv3}) except for the function $\mathcal{C}$. After the same discussion as in the previous subsection, one finally obtains the quadratic action for the vector perturbations: 
\begin{equation}\label{BV1}
	S^{(2)}_V= -\sum_{A=L,R} \int d\eta\, d^3k \, w_A^2 E^A E^{*A}~,
\end{equation}
where
\begin{equation}\label{pc1}
w_A^2= \frac{k^4}{a^2}c_4^2 \bar{\phi}^{\prime4} +\frac{1}{2}\lambda^A k^3 \left[\left(b_2-c_4\right) \bar{\phi}^{\prime2}\right]'.
\end{equation}
It is evidently an action for non-dynamical fields, and free of ghost instability of course. 

So far we have derived the whole linear perturbation theory on the non-degenerate backgrounds (\ref{Sol1}). We found that the results are the same as ones on the degenerate background family(\ref{Sol3}).

\subsection{Quadratic actions on the non-degenerate backgrounds 2}
\subsubsection{Scalar perturbations}
For scalar perturbations, same as the cases of degenerate backgrounds and non-degenerate backgrounds 1, the parity violating terms have no contribution to the quadratic action, so the quadratic action for the scalar perturbations of the model (\ref{S2}) is also the same as the one in GR with a minimally coupled scalar field:
\begin{equation}
	S^{(2)}_S= \int d^4x \frac{a^2\bar{\phi}^{\prime 2}}{2\mathcal{H}^2}\left(\zeta'^2-\partial_i\zeta\partial_i\zeta\right)~,
\end{equation}
where $\zeta=-(\psi+\mathcal{H}\delta\phi/\phi')$ is the Mukkhanov-Sasaki variable. Similarly, this quadratic action is free of effect of the perturbations $u^0$ and $u$ from the connection.

\subsubsection{Tensor perturbations}
The tensor perturbations on this family of backgrounds (\ref{Sol2}) behave differently from the rest two families of backgrounds. In this case, $M_4$ also has contribution together with $M_1$ and $M_5$:
\begin{equation}
    \begin{aligned}
        S_{PV1}&=\int d^4x\,2c_1\,\bar{\phi}^{\prime 2} \epsilon_{ijk}\left(h_{jl,i}h'_{kl} +Fh_{jl,i}h_{kl}\right),\\
        S_{PV4}&=\int d^4x\,c_4\,\bar{\phi}^{\prime 2} \epsilon_{ijk} F h_{jl,i}h_{kl},\\
        S_{PV5}&=-\int d^4x\,c_5\,\bar{\phi}^{\prime 2} \epsilon_{ijk}h_{jl,i}h'_{kl}.
    \end{aligned}
\end{equation}
Therefore, different from (\ref{qSt3}) and (\ref{qSt1}), the quadratic action for tensor perturbations on backgrounds (\ref{Sol2}) is
\begin{equation}
\begin{aligned}
    S^{(2)}_T= \int d^4x &\left\lbrace \frac{a^2}{8}\left( h'_{ij}h'_{ij} -h_{ij,k} h_{ij,k}\right)\right.\\
    &~~~\left.+\left[ (2c_1+c_4)F\bar{\phi}^{\prime 2}-\frac{1}{2} \left(b_2\bar{\phi}^{\prime 2}\right)' \right] \epsilon_{ijk}h_{jl,i}h_{kl}\right\rbrace,
\end{aligned}
\end{equation}

The effective action in Fourier space still has a similar form:
\begin{equation}\label{qStt2}
	S^{(2)}_T=\sum_{A=L,R} \int d\eta\, d^3\vec{k}\,\frac{a^2}{8} \left(h'^A h'^{A*}-\omega^{2}_{AT}h^A h^{A*}\right)~,
\end{equation}
but is with a different dispersion relation:
\begin{equation}\label{drT2}
	\omega^2_{AT}=k^2\left[ 1+\frac{4\lambda_A}{a^2k} \left(b_2\bar{\phi}^{\prime2}\right)' -\frac{8\lambda^A F}{a^2 k}(2c_1+c_4)\bar{\phi}^{\prime 2}\right]~.
\end{equation}
So we have a qualitatively similar but quantitatively different conclusion: the quadratic action (\ref{qStt2}) shows no ghost mode in the tensor perturbations, for both of the two modes of GWs have the right sign ($a^2>0$) in their kinetic terms; while the helicity dependent dispersion relation (\ref{drT2}) implies the velocity birefringence phenomenon. Furthermore, the dispersion relation (\ref{drT2}) is quantitatively different from ones on the degenerate backgrounds (\ref{drT3}) and the non-degenerate backgrounds 1 (\ref{drT1}): an additional term dependent on function $F$ is included.

\subsubsection{Vector perturbations}
After some calculations, one can obtain the parity violating terms that contribute to the quadratic action for the vector perturbations: 
\begin{equation}
\begin{aligned}
	S_{PV1}&=\int d^4x\, 2c_1 \bar{\phi}^{\prime2} \epsilon_{ijk}\left( B_{j,i}B'_k +\frac{F'}{F} B_{j,i}B_k -E_{j,il}E'_{k,l} +FE_{j,il}E_{k,l} \right),\\
	S_{PV2}&=\int d^4x\, 2c_2 \bar{\phi}^{\prime2} \epsilon_{ijk}\left[B_{j,i}B'_k +\left(\frac{F'}{F}+F\right)B_{j,i}B_k\right], \\
	S_{PV4}&=-\int d^4x\, c_4 \bar{\phi}^{\prime2} \epsilon_{ijk}\left( B_{j,i}B'_k +\frac{F'}{F}B_{j,i}B_k -E_{j,il}B_{k,l} +FE_{j,il}E_{k,l}\right), \\
	S_{PV5}&=\int d^4x\, c_5 \bar{\phi}^{\prime2} \epsilon_{ijk}\left[E_{j,il}E'_{k,l} -B_{j,i}B'_k -\left(\frac{F'}{F}-F\right) B_{j,i}B_k\right],
\end{aligned}
\end{equation}
Different from the cases in the last two subsections, even if $b_1=0$, the function $F$ still influence the quadratic action. Summing terms above with vector sector of $S_{STEGR}$, we have the full quadratic action for the vector perturbations of the model (\ref{S2}): 
\begin{equation}\label{qSv2}
\begin{aligned}
    S^{(2)}_V=\int d^4x\, &\left\lbrace \frac{1}{4}a^2(B_{i,j}+E'_{i,j})(B_{i,j}+E'_{i,j}) +\epsilon_{ijk}\bar{\phi}^{\prime2}\left[ b_1  B_{j,i}B'_k +b_1\frac{F'}{F} B_{j,i}B_k +\left(2c_2+c_5\right)FB_{j,i}B_k\right.\right.\\
    &\quad\left.\left. +c_4 E_{j,il}B_{k,l} -b_2 E_{j,il}E'_{k,l} +\left(2c_1+c_4\right)FE_{j,il}E_{k,l}\right] \right\rbrace~.
\end{aligned}	
\end{equation}
Set $b_1=0$ and conduct the same discussion as in the previous subsections, one finally obtains the quadratic action for the vector perturbations:
\begin{equation}\label{BV2}
	S^{(2)}_V= \sum_{A=L,R} \int d\eta d^3k \, \left( z_A^2 E'^A E'^{*A} -w_A^2 E^A E^{*A}\right)~,
\end{equation}
where the coefficients of kinetic terms
\begin{equation}\label{kc2}
	z_A^2= \frac{\lambda^A \left(2c_2+c_5\right)F\bar{\phi}^{\prime2} a^2 k^2}{2a^2 k+4\lambda^A \left(2c_2+c_5\right)F\bar{\phi}^{\prime2}}~.
\end{equation}
Coefficients $z_A^2$ implies ghost instability in vector modes. When $\left(2c_2+c_5\right)\bar{\phi}^{\prime2}\neq0$ (since $F\neq0$), the two polarization modes of vector perturbations are both dynamical. At small length scales, where $k\gg 2 a^{-2}\left| \left(2c_2+c_5\right)\bar{\phi}^{\prime2}\right| $, $z_A^2\simeq\lambda^A \left(2c_2+c_5\right)\bar{\phi}^{\prime2} k/2$. Since $\lambda^A=\mp1$, one of the two polarization modes has a negative sign in kinetic term and becomes a ghost mode. Therefore, condition $b_1=0$ is not enough to cancel the ghost modes on the non-degenerate family 2 of backgrounds (\ref{Sol2}).

In order to cancel the ghost modes on backgrounds (\ref{Sol2}), an additional condition $\left(2c_2+c_5\right)\bar{\phi}^{\prime2}=0$ is necessary, which gives us two choices: $\bar{\phi}'=0$ or $2c_2+c_5=0$. The former requires that $\bar{\phi}=\text{const}$ on the background, which is difficult to satisfy. So we may choose the latter $2c_2+c_5=0$ to cancel the ghost modes. Under this condition, the quadratic action for vector perturbations finally reduces to
\begin{equation}\label{BV2r}
	S^{(2)}_V= -\sum_{A=L,R} \int d\eta\, d^3k \, w_A^2 E^A E^{*A}~,
\end{equation}
where
\begin{equation}\label{pc2}
w_A^2= \frac{k^4}{a^2}c_4^2 \bar{\phi}^{\prime4} +\frac{1}{2}\lambda^A k^3 \left[\left(b_2-c_4\right) \bar{\phi}^{\prime2}\right]' -\frac{1}{2}\lambda^A k^3\left(2c_1-c_4\right) \bar{\phi}^{\prime2}.
\end{equation}
It is evidently an action for non-dynamical fields, and free of ghost instability. Compared with ones on other backgrounds (\ref{pc3},\ref{pc1}), in (\ref{pc2}) an additional term dependent on function $F$ is included.

\section{Conclusion}\label{conclusion}

In this paper, we further studied the STG model (\ref{S2}) proposed in our previous work, which modifies QGR (\ref{Qac}) by several parity violating interactions between the gravitational field and a scalar field. We introduced three families of background solutions in flat FRW universe, which still satisfy the cosmological principle, and pointed out that the simplest background solution considered in our previous work is a special case in the degenerate family (\ref{Sol3}).

We then discussed the linear perturbation theory on these three families of backgrounds. Our results showed that, background dynamics and the behavior of scalar perturbations remain the same as GR coupled with a scalar field. The tensor perturbations present velocity birefringence phenomenon in GWs and are free of ghost instability, while the dispersion relation on the non-degenerate background family 2 (\ref{Sol2}) is quantitatively different from ones on the other two families.

After attaching the ghost-free condition $b_{1}\equiv 2c_1+2c_2-c_4-c_5=0$ in our previous work to the calculations, we found that on the degenerate family (\ref{Sol3}) and non-degenerate family 1 (\ref{Sol1}) of background, the vector perturbations do not propagate and thus are not ghost modes. On the other hand, on the non-degenerate background family 2 (\ref{Sol2}), the vector perturbations are promoted to be dynamical propagating fields, present both amplitude and velocity birefringence phenomena, and still suffer from ghost instability, generally. In order to avoid ghost instability, an additional condition $2c_2+c_5=0$ should also be satisfied. However, if we believed the non-degenerate solutions family 2 (\ref{Sol2}) was not physically permissible, the condition $2c_2+c_5=0$ would not be necessary.

We also notice that although we remove two parity violating terms through the two identities (\ref{PVId}), there is still one term vanishing up to the second order for the FRW background and have no contribution to the quadratic actions. We believe it merits further analysis whether the term has effect for linear perturbations around other backgrounds.

\section{Acknowledgments}
This work is supported by the National Key R\&D Program of China Grant No. 2021YFC2203102 and by NSFC under Grant No. 12075231 and 12247103.

\appendix
\section{Perturbation Expansion of Connection in Symmetric Teleparallel Framework}

The background and perturbed connections are
\begin{align}
        \bar{\Gamma}^\lambda_{~~\mu\nu}&=\frac{\partial x^\lambda}{\partial \bar{y}^\alpha}\partial_\mu \partial_\nu \bar{y}^\alpha,\label{bkcon}\\
        {\Gamma^\lambda}_{\mu\nu}&=\frac{\partial x^\lambda}{\partial y^\alpha}\partial_\mu \partial_\nu y^\alpha,\label{pcon}
\end{align}
respectively, where $y^\alpha=\bar{y}^\alpha+\delta y^\alpha$. From (\ref{bkcon}), One would obtain $\bar{\Gamma}^\lambda_{~~\mu\nu}\partial_\lambda \bar{y}^\alpha =\partial_\mu \partial_\nu \bar{y}^\alpha$.

Noticing that the derivative of inverse of a matrix is $d M^{-1}=-\left(M^{-1}\right)^T~dM~M^{-1}$, we have
\begin{equation}
    \begin{aligned}
        \partial_\mu \left(\frac{\partial x^\lambda}{\partial \bar{y}^\alpha}\right) &=\partial_\mu \left[\left(\frac{\partial \bar{y}^\alpha}{\partial x^\lambda}\right)^{-1}\right]\\
        &=-\frac{\partial x^\nu}{\partial \bar{y}^\alpha} \frac{\partial x^\lambda}{\partial \bar{y}^\beta} \partial_\mu \partial_\nu \bar{y}^\beta\\
        &=-\frac{\partial x^\nu}{\partial \bar{y}^\alpha} \bar{\Gamma}^\lambda_{~~\mu\nu}.
    \end{aligned}
\end{equation}
Marking
\begin{equation}
    u^\lambda =\delta y^\alpha \frac{\partial x^\lambda}{\partial \bar{y}^\alpha},
\end{equation}
then we have
\begin{equation}
    \begin{aligned}
        \partial_\mu u^\lambda &=\partial_\mu \left(\delta y^\alpha \frac{\partial x^\lambda}{\partial \bar{y}^\alpha}\right)\\
        &=\frac{\partial x^\lambda}{\partial \bar{y}^\alpha} \partial_\mu \delta y^\alpha +\partial_\mu \left(\frac{\partial x^\lambda}{\partial \bar{y}^\alpha}\right) \delta y^\alpha\\
        &=\frac{\partial x^\lambda}{\partial \bar{y}^\alpha} \partial_\mu \delta y^\alpha -\frac{\partial x^\nu}{\partial \bar{y}^\alpha} \delta y^\alpha \bar{\Gamma}^\lambda_{~~\mu\nu},
    \end{aligned}
\end{equation}
which yields
\begin{equation}
    \frac{\partial x^\lambda}{\partial \bar{y}^\alpha} \partial_\nu \delta y^\alpha =\bar{\nabla}_\nu u^\lambda,~\partial_\nu \delta y^\alpha =\partial_\lambda \bar{y}^\alpha \bar{\nabla}_\nu u^\lambda,
\end{equation}
and
\begin{equation}\label{in1}
        \frac{\partial \delta y^\beta}{\partial \bar{y}^\alpha} =\frac{\partial x^\sigma}{\partial \bar{y}^\alpha} \partial_\sigma \delta y^\beta = \frac{\partial x^\sigma}{\partial \bar{y}^\alpha} \partial_\lambda \bar{y}^\beta \bar{\nabla}_\sigma u^\lambda;
\end{equation}
furthermore, the higher derivative of $\delta y^\alpha$ is
\begin{equation}
    \begin{aligned}
        \partial_\mu \partial_\nu \delta y^\alpha &=\partial_\mu \left(\partial_\lambda \bar{y}^\alpha \bar{\nabla}_\nu u^\lambda\right)\\
        &=\partial_\mu \partial_\lambda \bar{y}^\alpha \bar{\nabla}_\nu u^\lambda +\partial_\lambda \bar{y}^\alpha \partial_\mu \left(\bar{\nabla}_\nu u^\lambda\right)\\
        &=\partial_\lambda \bar{y}^\alpha \partial_\mu \left(\bar{\nabla}_\nu u^\lambda\right) +\partial_\lambda \bar{y}^\alpha \bar{\Gamma}^\lambda_{~~\mu\rho}\bar{\nabla}_\nu u^\rho\\
        &=\partial_\lambda \bar{y}^\alpha \left(\bar{\nabla}_\mu \bar{\nabla}_\nu u^\lambda +\bar{\Gamma}^\sigma_{~~\mu\nu}\bar{\nabla}_\sigma u^\lambda\right),
    \end{aligned}
\end{equation}
thus
\begin{equation}\label{o1}
    \bar{\nabla}_\mu \bar{\nabla}_\nu u^\lambda =\frac{\partial x^\lambda}{\partial \bar{y}^\alpha}\partial_\mu \partial_\nu \delta y^\alpha -\frac{\partial \delta y^\beta}{\partial \bar{y}^\alpha} \frac{\partial x^\lambda}{\partial \bar{y}^\beta}\partial_\mu \partial_\nu \bar{y}^\alpha;
\end{equation}
and with (\ref{in1}) we have
\begin{equation}\label{o2}
    \bar{\nabla}_\mu \bar{\nabla}_\nu u^\rho \bar{\nabla}_\rho u^\lambda =\frac{\partial \delta y^\beta}{\partial \bar{y}^\alpha} \frac{\partial x^\lambda}{\partial \bar{y}^\beta}\partial_\mu \partial_\nu \delta y^\alpha -\frac{\partial \delta y^\beta}{\partial \bar{y}^\gamma} \frac{\partial \delta y^\gamma}{\partial \bar{y}^\alpha} \frac{\partial x^\lambda}{\partial \bar{y}^\beta}\partial_\mu \partial_\nu \bar{y}^\alpha. 
\end{equation}

Now derive the perturbation expansion of Eq.(\ref{pcon}). From the expansion of the inverse of a matrix, one would obtain
\begin{equation}
    \begin{aligned}
        \frac{\partial x^\lambda}{\partial y^\alpha}&=\frac{\partial x^\lambda}{\partial \bar{y}^\beta}\frac{\partial \bar{y}^\beta}{\partial y^\alpha}\\
        &=\frac{\partial x^\lambda}{\partial \bar{y}^\beta}\left(\delta^\alpha_\beta +\frac{\partial \delta y^\alpha}{\partial \bar{y}^\beta}\right)^{-1}\\
        &=\frac{\partial x^\lambda}{\partial \bar{y}^\beta}\left(\delta^\beta_\alpha -\frac{\partial \delta y^\beta}{\partial \bar{y}^\alpha} + \frac{\partial \delta y^\beta}{\partial \bar{y}^\gamma}\frac{\partial \delta y^\gamma}{\partial \bar{y}^\alpha}+\cdots\right).
    \end{aligned}
\end{equation}
Substitute it into Eq.(\ref{pcon}): 
\begin{equation}\label{pcon2}
    \begin{aligned}
        {\Gamma^\lambda}_{\mu\nu}&=\frac{\partial x^\lambda}{\partial \bar{y}^\alpha}\partial_\mu \partial_\nu \bar{y}^\alpha +\frac{\partial x^\lambda}{\partial \bar{y}^\alpha}\partial_\mu \partial_\nu \delta y^\alpha -\frac{\partial x^\lambda}{\partial \bar{y}^\beta}\frac{\partial \delta y^\beta}{\partial \bar{y}^\alpha}\partial_\mu \partial_\nu \bar{y}^\alpha\\ &\quad+\frac{\partial x^\lambda}{\partial \bar{y}^\beta}\frac{\partial \delta y^\beta}{\partial \bar{y}^\gamma}\frac{\partial \delta y^\gamma}{\partial \bar{y}^\alpha}\partial_\mu \partial_\nu \bar{y}^\alpha -\frac{\partial x^\lambda}{\partial \bar{y}^\beta}\frac{\partial \delta y^\beta}{\partial \bar{y}^\alpha}\partial_\mu \partial_\nu \delta y^\alpha +\cdots.
    \end{aligned}
\end{equation}
Then using Eq.(\ref{o1},\ref{o2}) to simplify the formula above, finally we have
\begin{equation}
        {\Gamma^\lambda}_{\mu\nu}=\bar{\Gamma}^\lambda_{~~\mu\nu} +\bar{\nabla}_\mu \bar{\nabla}_\nu u^\lambda -\bar{\nabla}_\mu \bar{\nabla}_\nu u^\rho \bar{\nabla}_\rho u^\lambda +\cdots,
\end{equation}
which is just Eq.(\ref{percon}).

\end{document}